\pdfoutput=1
\PassOptionsToPackage{table, xcdraw}{xcolor}
\documentclass[11pt]{article}

\usepackage[final]{acl}
\usepackage{algorithm}
\usepackage{algpseudocode}
\usepackage{amssymb}
\usepackage{makecell}
\usepackage{hyperref}
\usepackage{times}
\usepackage{latexsym}
\usepackage{amssymb}
\usepackage{booktabs}

\usepackage[table, xcdraw]{xcolor}

\usepackage{colortbl} 
\usepackage{multirow}
\usepackage[T1]{fontenc}

\usepackage[utf8]{inputenc}

\usepackage{microtype}
\usepackage{amsmath}
\algblock{Output}{EndOutput}
\algnotext{EndOutput}
\DeclareMathOperator{\softmax}{\mathrm{Softmax}}

\DeclareMathOperator{\append}{\mathrm{append}}
\usepackage{inconsolata}

\usepackage{graphicx}

\title{\raisebox{-0.22\height}{\includegraphics*[width=1.2cm]{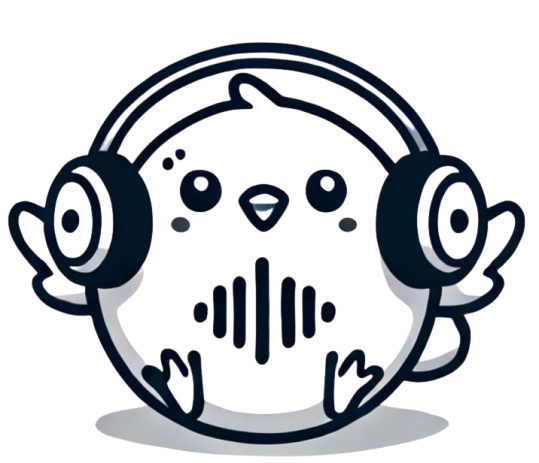}}~FastAdaSP: Multitask-Adapted Efficient Inference for Large Speech Language Model}

\author{
 \textbf{Yichen Lu\textsuperscript{1}\thanks{Equal Contributions.}} \quad
 \textbf{Jiaqi Song\textsuperscript{1}\footnote[1]{}} \quad
 \textbf{Chao-Han Huck Yang\textsuperscript{2}} \quad
 \textbf{Shinji Watanabe\textsuperscript{1}}
\\
 \textsuperscript{1}Carnegie Mellon University \quad
 \textsuperscript{2}NVIDIA Research
\\
\texttt{\{yichenl5,~jiaqison,~swatanab\}@andrew.cmu.edu \quad hucky@nvidia.com} 
\\
}

\begin{document}
\maketitle
\begin{abstract}
In this study, we aim to explore Multitask Speech Language Model (SpeechLM) efficient inference via token reduction. Unlike other modalities such as vision or text, speech has unique temporal dependencies, making previous efficient inference works on other modalities not directly applicable. Furthermore, methods for efficient SpeechLM inference on long sequence and sparse signals remain largely unexplored. Then we propose \textbf{FastAdaSP}, a weighted token merging framework specifically designed for various speech-related tasks to improve the trade-off between efficiency and performance. Experimental results on WavLLM and Qwen-Audio show that our method achieves the state-of-the-art (SOTA) efficiency-performance trade-off compared with other baseline methods. Specifically, FastAdaSP achieved \textbf{7x} memory efficiency and \textbf{1.83x} decoding throughput without any degradation on tasks like Emotion Recognition (ER) and Spoken Question Answering (SQA). The code will be available at \url{https://github.com/yichen14/FastAdaSP}
\end{abstract}

\section{Introduction}

Speech Language Models (SpeechLMs) have been an important role in the field of natural language processing and speech technology. Recent advancements~\cite{hu2024wavllm, chu2023qwenaudio, tang2024salmonn}~have demonstrated significant capabilities in voice processing and audio understanding. Furthermore, GPT4-o~\cite{openai2024gpt4technicalreport} showcases conversational speech processing abilities, advancing the capability of LLMs toward various voice-interface applications.
However, challenges related to inference latency and memory efficiency remain major bottlenecks, especially as multitask SpeechLMs grow larger, reaching up to 7 billion parameters. These challenges necessitate the development of more efficient inference methods.


\begin{figure}[htbp]
\centering
\includegraphics[width=0.42\textwidth]{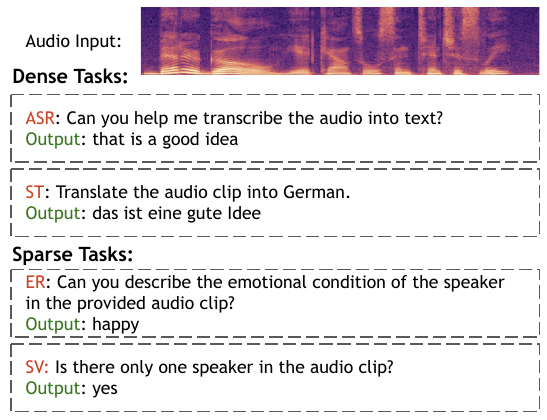} \\
\caption{Examples of \textbf{Multitask SpeechLM} on dense (ASR, ST) and sparse (ER, SV) tasks}
\label{fig:exp}
\end{figure}

SpeechLMs are often capable of performing a wide range of speech or audio-related tasks. As shown in Figure \ref{fig:exp}, in our study, we categorize and define these tasks into two distinct classes: \textbf{Dense Tasks:} Nearly all input audio tokens are useful, such as in Automatic Speech Recognition (ASR) and Speech Translation (ST); \textbf{Sparse Tasks:} Tasks like Emotion Recognition (ER) and Speaker Verification (SV), where only a few tokens within the entire audio input contain the crucial information needed to perform the task.




The temporal dependencies in speech signals require efficient handling of long sequences, while the sparsity of relevant information demands precise extraction of crucial audio features. These unique properties make SpeechLM tasks distinct from other modalities like vision or text, especially when implementing token reduction techniques.

To address these issues and improve the efficiency of SpeechLM inference, we introduce \textbf{FastAdaSP}, a unified SpeechLM fast inference framework that incorporates multiple audio token reduction methods during the pre-filling stage tailored to different types of tasks. FastAdaSP does not require any additional training, making the entire framework more practical and easy to use. Our main contributions are as follows:

1. We introduce a new plug-and-play method for effectively selecting layers for audio token reduction operations on sparse tasks.

2. We study efficient inference methods specifically designed for both dense and sparse tasks on SpeechLMs and validate the effectiveness of our methods across multiple tasks. 

3. To benchmark the task, previous token reduction methods, started from other modalities, have been investigated and analyzed in this emerging context of SpeechLM settings.

\section{Related Work}
\label{relatedword}
%
\textbf{Large Speech Language Models: }SpeechLMs ~\cite{borsos2023audiolm, rubenstein2023audiopalm, radhakrishnan2023whispering, tang2024salmonn, chu2023qwenaudio, hu2024wavllm, gong2024listenthinkunderstand, maiti2024voxtlm,lu2024syneslmunifiedapproachaudiovisual} adopt a large pretrained language model ~\cite{touvron2023llamaopenefficientfoundation} as their base model and use audio encoder(s) ~\cite{radford2022robustspeechrecognitionlargescale, wavlm, hsu2021hubertselfsupervisedspeechrepresentation} to process raw audio input. Leveraging the language understanding and reasoning abilities of LLMs, SpeechLMs can perform various speech-related tasks. However, as SpeechLMs grow in size, inference latency and memory efficiency become problematic. Thus, research on cost-saving techniques is essential to address these challenges.



\noindent\textbf{Efficient Inference in ASR:} Recent studies~\cite{zhu2024skipformerskipandrecoverstrategyefficient, kim2022squeezeformerefficienttransformerautomatic, efficient-conf} have focused on efficient inference for ASR models~\cite{conformer, branchformer} by progressively down-sampling the audio features in the audio encoder to reduce sequence length. However, these methods are specifically designed for the ASR task and do not generalize well to multitask settings for SpeechLMs.


\noindent\textbf{Key-Value (KV) Cache Compression:} 
In addition to the efficient inference methods for ASR, some of other works are focusing on compressing KV Cache to speed-up LLMs inference. Previous works such as StreamLLM~\cite{xiao2023efficient}, H$_2$O~\cite{zhang2023h2o}, LESS~\cite{dong2024less}, LOOK-M~\cite{wan2024lookmlookonceoptimizationkv} were designed to compress the text or vision KV cache during inference to overcome the limited KV cache size and accelerate the inference speed. However, KV cache compression techniques do not actually reduce the number of input tokens during the pre-filling stage. When a long video is input to a multimodal LLM, the extensive sequence of vision and audio tokens can exceed the context length limit of the backbone LLM, causing several issues. Moreover, this technique does not improve the latency of the pre-filling stage.




\noindent\textbf{Token Reduction:} To address these issues, extensive research has been conducted on token pruning techniques within Vision Language Models (VLMs). Recently, lots of token reduction works such as FastV~\cite{chen2024imageworth12tokens}, ToMe~\cite{bolya2023token}, LLava-PruneMerge~\cite{shang2024llavaprumergeadaptivetokenreduction} focus on reducing the vision tokens to lower the computational costs through token eviction or merge. Besides the vision modality, A-ToMe~\cite{li2023accelerating} applied the ToMe~\cite{bolya2023token} method to the audio modality in a Transformer-transducer model~\cite{transducer} for ASR tasks only. However, token reduction methods for the audio modality in multitask SpeechLMs remain unexplored. Inspired by these previous works, our study primarily develops token reduction techniques that combine token merging and eviction for the audio modality in SpeechLMs during the inference process. We also explore the applicability of these methods to various speech-related tasks.

\section{Methodology}


\begin{figure*}[ht!]
\centering
\includegraphics[width=0.99\textwidth]{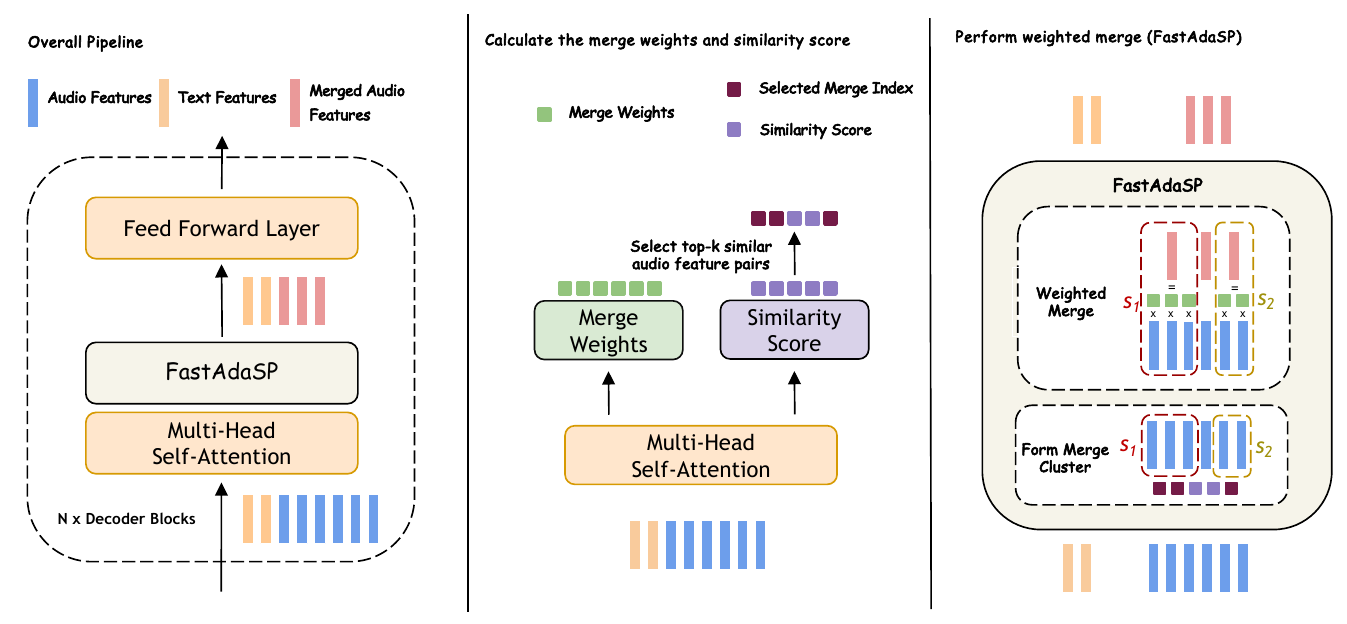} \\
\caption{\textbf{FastAdaSP: Weighted Token Merge} of audio features in the decoder blocks of SpeechLMs  }
\label{fig:model}
\end{figure*}

In this section, we introduce the motivation and formulation of FastAdaSP, followed by our layer selection and task-specific design strategies for Multitask SpeechLMs. Note that, in our work, \textit{audio tokens} refers to the audio features output by the multi-head attention block.

\subsection{Preliminary}

\noindent\textbf{Speech Modality in Multitask SpeechLMs:} During inference, VLMs often use only a \textit{small portion} of visual information for reasoning and context understanding. However, SpeechLMs are capable of performing multiple tasks within a single model. For sequence-to-sequence dense tasks like ASR, it is crucial to consider ``all audio tokens'' to generate accurate transcriptions. In addition to dense tasks, SpeechLMs also need to perform sparse tasks such as ER and SQA, where only a few tokens in the input hold critical information for generating accurate predictions. Therefore, a more careful token reduction policy is necessary for SpeechLMs.

\noindent\textbf{Pre-filling Phase of SpeechLMs:} During the pre-filling phase of SpeechLMs, the raw audio sequence is usually processed by pre-trained audio encoder(s) to extract the semantic and acoustic information into the embedding space $\mathbf{X}_{\text{audio}} \in \mathbb{R}^{L_{\text{audio}} \times D}$. Consider the text embedding of user instruction $\mathbf{X}_{\text{text}} \in \mathbb{R}^{L_{\text{text}} \times D}$, the input to the decoder blocks of SpeechLM is $\mathbf{X} \in \mathbb{R}^{L_{\text{prompt}} \times D}$, which represented as $\mathbf{X} = [\mathbf{X}_{\text{audio}}, \mathbf{X}_{\text{text}}]$. Here, $L_\text{prompt}$ is the sum of audio embedding length $L_\text{audio}$ and text embedding length $L_\text{text}$, and $D$ is the model's hidden dimension.

In each self-attention block of the transfomer decoder layer, the query, key, value tensors can be derived by:
\begin{equation}
    \mathbf{Q} = \mathbf{X}\mathbf{W}_Q,~\mathbf{K} = \mathbf{X}\mathbf{W}_K,~\mathbf{V} = \mathbf{X}\mathbf{W}_V~,
\end{equation}

where $\mathbf{W}_Q,\mathbf{W}_K,\mathbf{W}_V \in \mathbb{R}^{D \times D}$ represents the matrix weights for query, key, and value layers, respectively. After this computation, the value of $\mathbf{K}, \mathbf{V}$ will be stored in the KV cache which will be used in the decoding phase. Then the self-attention output can be computed as:
\begin{equation}
    \label{attn}
    \mathbf{X}_{\text{attention}} = \softmax \left( \frac{\mathbf{Q} \mathbf{K}^\top}{\sqrt{D}} \right)\mathbf{V}.
\end{equation}

\noindent\textbf{Decoding Phase of SpeechLMs:} During the auto-regressive decoding phase of SpeechLMs, the KV cache is employed and updated for all the new generated tokens. At each step, the total key and value are calculated by using the previous stored $\mathbf{K}_\text{cache}$ and $\mathbf{V}_\text{cache}$ and the new input $\mathbf{X_\text{new}}$ as:
\begin{equation}
    \mathbf{K} = [\mathbf{K}_\text{cache},\mathbf{X_\text{new}}\mathbf{W}_K], \mathbf{V} = [\mathbf{V}_\text{cache},\mathbf{X_\text{new}}\mathbf{W}_V].
\end{equation}

Equation \ref{attn} is used to calculate the attention output. During this stage, the KV cache grows linearly, and each new token significantly increases memory consumption and attention computation latency, especially when the generated sequence is very long.

\subsection{FastAdaSP: Method}
\label{method}
To accommodate both sparse and dense tasks in SpeechLMs, we designed a novel token reduction method with different strategies for each.

\noindent\textbf{Weighted Token Merge:} Dense tasks like ASR require most of the token information during inference, making direct token dropping from the attention output too aggressive and likely to result in the loss of critical information. Instead, merging similar audio tokens can eliminate redundant audio information while preserving essential content. 

Token merge techniques in the vision modality require calculating the similarity between numerous pairs of image patches in the spatial domain to identify the most similar pairs for merging~\cite{bolya2023token}. For audio signals, however, token merge in audio processing needs to operate in the temporal domain. This involves calculating the similarity along adjacent audio tokens pairs and merge a \textit{cluster} of adjacent audio tokens for a sequence of audio features $A = (\mathbf{a}_{i}\in\mathbb{R}^D|i = 1, ..., L)$. For the audio features from $1$ to $L-1$, we use the cosine similarity score between the adjacent audio token key state to determine their similarity:
\begin{equation}
    p_i = \frac{\mathbf{K}_i^\top \cdot \mathbf{K}_{i+1}}{\|\mathbf{K}_i\| \|\mathbf{K}_{i+1}\|}
\end{equation}


We then obtain an adjacent similarity score sequence $P = (p_i \in \mathbb{R} | i = 1, ..., L-1)$. After determining the number of tokens to merge, we select the top-k largest adjacent similarity indices to form the merge index list. Next, we loop through the merge index list, grouping multiple adjacent indices into a single merge cluster. Finally, we obtain $m$ merging clusters $S = \{s_i | i = 1, ..., m\}$ where $s_i$ represent a merging cluster which contains several adjacent audio tokens. Then we obtain the merge weights $W_\text{merge}=(\omega_{i}\in\mathbb{R}|i = 1, ..., L)$ for audio features $\mathbf{A}$ by: 
\begin{equation}
    \omega_i = (\sum^{H}\sum^{L_\text{prompt}} \softmax \left( \frac{\mathbf{Q} \mathbf{K}^\top}{\sqrt{D}} \right))_i
\end{equation}
\textbf{}
Where $L_\text{prompt} = L_{\text{audio}} + L_{\text{text}}$ represents the query length. $H$ represents the number of attention head. Both audio and text features are utilized to calculate the overall cumulative attention score. By leveraging the interaction between text instructions and speech, we can determine the importance of audio tokens in the current context. The merged audio feature $\mathbf{a}_i^{\text{merge}}$ for each cluster $s_i$ will be calculated as: 
\begin{equation}
    \mathbf{a}_i^{\text{merge}} = \frac{\sum^{|s_i|}\omega_j\mathbf{a}_j}{\sum^{|s_i|}\omega_j}
\end{equation}

The overall procedure of the weighted token merge is shown in Figure \ref{fig:model}. This method selects the relatively important tokens to keep and the redundant tokens to drop at that layer, effectively preserving as much information as possible while significantly reducing the number of tokens. For full details of the algorithm, please refer to \ref{algo}.

\begin{table*}[]
\centering
\resizebox{\textwidth}{!}{
\begin{tabular}{@{}c|cccc@{}}
\toprule
Model                      & Task                                   & Dataset   & Split       & Metric \\ \midrule
\multirow{4}{*}{WavLLM~\cite{hu2024wavllm}}    & Automatic Speech Recognition (ASR)     & LibriSpeech~\cite{7178964} & dev | test & WER    \\
                           & Speech Translation (ST)                & Must-C~\cite{di-gangi-etal-2019-must}   & en-de        & BLEU   \\
                           & Emotion Recognition (ER)               & IEMOCAP~\cite{busso2008iemocap}    &  Session 5     & ACC    \\
                           & Spoken Language Answering (SQA)        & MuTual~\cite{cui2020mutualdatasetmultiturndialogue}   & test        & ACC    \\ \midrule
\multirow{4}{*}{Qwen-Audio~\cite{chu2023qwenaudio}} & Automatic Speech Recognition (ASR) & LibriSpeech~\cite{7178964} &  dev | test & WER    \\
                           & Speech Translation (ST)                & CoVoST2~\cite{wang2020covost2massivelymultilingual}    & en-zh      & BLEU   \\
                           & Emotion Recognition (ER)               & MELD~\cite{poria2019meldmultimodalmultipartydataset}    & test         & ACC    \\
                           & Audio Caption (AC)                     & Clotho~\cite{drossos2019clothoaudiocaptioningdataset}   & test        & CIDEr | SPICE | SPIDEr \\ \bottomrule
\end{tabular}}
\caption{\textbf{Task, dataset, and metrics} in the experiments}
\label{tasks}
\end{table*}
\begin{table*}[htbp]
\centering
\resizebox{\textwidth}{!}{\begin{tabular}{@{}c|ccccc|ccccc@{}}
\toprule
                                                                                            & \multicolumn{5}{c|}{ASR (WER\% $\downarrow$)}                                              & \multicolumn{5}{c}{ST (BLEU $\uparrow$)}                                                       \\ \midrule
\rowcolor[HTML]{C0C0C0} 
Full Token Baseline                                                                                    & \multicolumn{5}{c|}{\cellcolor[HTML]{C0C0C0}2.25}                             & \multicolumn{5}{c}{\cellcolor[HTML]{C0C0C0}21.56}                                  \\ \midrule
FLOPs Reduce                                                                                & 10\%          & 20\%          & 30\%          & 40\%          & 50\%          & 10\%           & 20\%           & 30\%           & 40\%           & 50\%           \\ \midrule
\rowcolor[HTML]{EFEFEF} 
H$_2$O~\cite{zhang2023h2o}                                                                                         & 2.25          & 2.46          & 3.37          & 5.60           & 10.20          & 20.42          & 20.12          & 19.99          & 18.64          & 17.36          \\
Random Merge                                                                                & 2.32          & 2.51          & 3.08          & 8.10           & 77.42         & 20.73          & 19.34          & 18.23          & 15.69          & 9.24           \\
Random Evict                                                                                & 2.53          & 4.34          & 9.23          & 34.80          & 172.52        & 20.34          & 19.03          & 17.36          & 14.10           & 8.59           \\
A-ToMe~\cite{li2023accelerating}                                                                                      & 2.35          & 2.92          & 4.43          & 15.87         & 50.46         & 20.53          & 19.42          & 17.29          & 13.42          & 8.51           \\
FastV~\cite{chen2024imageworth12tokens}                                                                                & 2.37          & 4.94          & 13.84         & 51.10          & 185.79        & 21.17          & 20.18          & 18.98          & 16.36          & 10.07          \\ \midrule
\textbf{\begin{tabular}[c]{@{}c@{}}\small{FastAdaSP-Dense (Decay)}\end{tabular}}    & \textbf{2.27} & 2.57          & 2.74          & 3.53          & 6.09          & 20.92          & 20.59          & 19.66          & 18.06          & 16.40           \\
\textbf{\begin{tabular}[c]{@{}c@{}}\small{FastAdaSP-Dense  (Constant)}\end{tabular}} & \textbf{2.27} & \textbf{2.49} & \textbf{2.48} & \textbf{2.96} & \textbf{4.73} & \textbf{21.47} & \textbf{20.72} & \textbf{19.81} & \textbf{18.54} & \textbf{17.45} \\ \bottomrule
\end{tabular}}
\caption{Comparison between FastAdaSP with other token reduction methods on WavLLM \textbf{dense tasks}}
\label{dense}
\end{table*}
\subsection{FastAdaSP: Strategies}
\label{strategy}
Based above method, we designed two similar but slightly different strategies for dense and sparse tasks to achieve better performance:

\noindent\textbf{Dense Task Strategy:} For dense tasks, we designed an operation scheduler that smoothly merges tokens layer by layer to prevent aggressive token dropping in SpeechLM. We implemented a constant schedule to maintain a consistent merge ratio and a decay schedule that linearly decreases the merge ratio to zero at the final layer. Please refer to \ref{appendix_schedule} for the ablation study of schedulers.

\noindent\textbf{Sparse Task Strategy:} For sparse tasks, a more aggressive token reduction method can be applied by merging tokens within a single layer. However, layer selection needs to be approached carefully as it significantly affects task performance. Therefore, we incorporate a \textit{Transfer Entropy}(TE)-based layer selection method (Section \ref{layer}) specially designed for sparse tasks. 

\subsection{Addtional Studies on Layer Selection}
\label{layer}
Recent token reduction works~\cite{chen2024imageworth12tokens, shang2024llavaprumergeadaptivetokenreduction, bolya2023token, li2023accelerating} often struggle with selecting appropriate layers for token reduction. Due to the difficulties in interpreting current auto-regressive transformer models, understanding the exact properties of different layers during inference is challenging. Consequently, previous works have relied on empirical studies to test various layers and reduction ratios. This approach is impractical and lacks generalization for actual deployment. Therefore, we aim to explore a justification to serve as a theoretical attempt of token reduction layer selection.



By definition, entropy can reflect the information carried out by each layer. Here, we take \( F \) as the feature output by the attention block which contains both audio and text features. Inspired by~\cite{NEURIPS2022_86e7ebb1,lin2024mlpgoodtransformerlearner}, we use the Gaussian distribution as the probability distribution to approximate the distribution of each channel in \( F \). Thus, the entropy measurement of a single layer \(H(F)\) can be defined as (for a more detail derivation, please refer to \ref{te_derivation}):
\begin{equation}
\label{eq_te}
H(F) \propto H_\sigma(F) = \sum_i \log [\sigma(F^i)]
\end{equation}



Here, we calculate the entropy of each layer by summing the logarithm of the standard deviation($\sigma$) of the each channels (audio tokens) in \( F \). To assess the impact of weighted merge on a specific layer's contribution to the final output distribution, we calculate the Transfer Entropy to measure the information difference at the final layer based on the operation layer of our method. We define Transfer Entropy ($\mathrm{TE}_i$.) for layer $i$. $\mathrm{TE}_i$ is equal to:
\begin{equation}
\left|H\left(\Phi\left(F_{\text {final }} ; \mathbb{W}_{\text {final }}\right)\right)-H\left(F_{\text {final }} \mid \Phi\left(F_i ; \mathbb{W}_i\right)\right)\right|
\end{equation}
where \(\Phi(\cdot; \cdot)\) represents the token reduction operation described in Section \ref{method}. It takes the layer feature $F$ and merge weights $\mathbb{W}$ as input and outputs the features after weighted token merge. Then TE$_i$ is the absolute difference between the final hidden states whether the token reduction operation is applied to layer \(i\). The smaller the $\text{TE}_i$, the less the final information loss caused by the operation on layer $i$. We also analyze the effectiveness of our TE-based layer selection method in Sec. \ref{result}.


\begin{table*}[ht]
\centering
\resizebox{\textwidth}{!}{\begin{tabular}{@{}c|ccccc|ccccc@{}}
\toprule
                           & \multicolumn{5}{c|}{ER (ACC\% $\uparrow$)}                                                    & \multicolumn{5}{c}{SQA (ACC\% $\uparrow$)}                                                    \\ \midrule
\rowcolor[HTML]{C0C0C0} 
Full Token Baseline                   & \multicolumn{5}{c|}{\cellcolor[HTML]{C0C0C0}72.80}                                  & \multicolumn{5}{c}{\cellcolor[HTML]{C0C0C0}67.60}                                  \\ \midrule
FLOPs Reduce               & 10\%           & 20\%           & 30\%           & 40\%           & 50\%           & 10\%           & 20\%           & 30\%           & 40\%           & 50\%          \\ \midrule
\rowcolor[HTML]{EFEFEF} 
H$_2$O~\cite{zhang2023h2o}                       & 72.32          & 72.60           & 73.73          & 72.11          & 72.36          & 67.10           & 68.4           & 68.00             & 67.55          & 65.40          \\
Random Merge               & 72.76          & 72.44          & 72.19          & 72.28          & 72.52          & 67.40           & 68.00             & 67.40           & 67.95          & 68.30          \\
Random Evict               & 73.08          & 71.71          & 72.44          & 72.03          & 72.28          & 67.65          & 67.35          & 68.35          & 67.80           & 67.50          \\
A-ToMe~\cite{li2023accelerating}                     & 72.84          & \textbf{72.68} & 72.2           & 71.23          & 69.54          & 67.05          & 67.15          & 65.75          & 63.45          & 62.60          \\
FastV~\cite{chen2024imageworth12tokens}               & 72.76          & 72.52          & 71.55          & 71.47          & 70.66          & 67.45          & 67.25          & \textbf{68.45} & 68.10           & 67.95         \\ \midrule
\rowcolor[HTML]{FFFFFF} 
\textbf{FastAdaSP-Sparse} & \textbf{73.16} & 72.60           & \textbf{73.73} & \textbf{73.65} & \textbf{73.65} & \textbf{67.65} & \textbf{68.05} & 67.45          & \textbf{68.45} & \textbf{68.70} \\ \bottomrule
\end{tabular}}
\caption{Comparison between FastAdaSP with other token reduction methods on WavLLM \textbf{sparse tasks}}
\label{sparse}
\end{table*}
\begin{table*}[ht]
\centering
\resizebox{\textwidth}{!}{\begin{tabular}{@{}c|c|cccc@{}}
\toprule
FLOPs Reduction \%  & Device                    & Real Time Factor $\downarrow$ & Pre-filling Latency (s) $\downarrow$ & Decoding Latency (s) $\downarrow$ & Throughput (token/s) $\uparrow$ \\ \midrule
0.00                          & \multirow{2}{*}{A100 80G} & 0.126                         & 6.72                                 & 23.55                             & 3.10                            \\
50.00                            &                           & 0.077                         & 6.48                                 & 11.89                             & 5.72                            \\ \midrule
0.00                          & \multirow{2}{*}{H100 80G} & 0.039                         & 1.13                                 & 8.39                              & 8.70                            \\
50.00                         &                           & 0.026                         & 0.96                                 & 5.42                              & 12.55                           \\ \bottomrule
\end{tabular}}
\caption{\textbf{Long Sequence Computational cost experiments} on a 240s audio sample with a batch size = 5 on WavLLM using one A100 80GB GPU and one H100 80GB GPU. For the full results, please refer to Appendix \ref{fullexp} }
\label{computation}
\end{table*}

\begin{table}[ht]
\centering
\begin{tabular}{@{}c|c@{}}
\toprule
Token Reduce \% & Max Batch Size (not OOM) \\ \midrule
Full Token Baseline               & 10                            \\  
50              & 70                           \\ \bottomrule
\end{tabular}
\caption{\textbf{Memory Saving Experiments:} Approximate maximum batch size under 50\% token reduction for WavLLM using a 240s audio sample on 1$\times$A100 80GB.}
\label{memory}
\end{table}
\begin{table}[ht]
\centering
\resizebox{0.48\textwidth}{!}{\begin{tabular}{@{}c|ccccccc@{}}
\toprule
FLOPs Reduce       & TE   & TE Rank & 10\%                                   & 20\%                         & 30\%                          & 40\%                         & 50\%                         \\ \midrule
Layer 2            & 2.20 & 4       & 54.78                                  & 54.30                         & 54.06                         & 52.91                        & 52.10                         \\
Layer 9            & 2.17 & 3       & \cellcolor[HTML]{FFFFFF}\textbf{55.51} & \cellcolor[HTML]{FFFFFF}54.30 & \cellcolor[HTML]{FFFFFF}53.61 & \cellcolor[HTML]{FFFFFF}53.30 & \cellcolor[HTML]{FFFFFF}51.50 \\
Layer 12           & 2.29 & 5       & 54.75                                  & 53.96                        & 53.44                         & 52.72                        & 48.35                        \\
Layer 15           & 2.11 & 2       & 53.98                                  & 54.06                        & 53.02                         & 50.57                        & -                            \\ \midrule
\textbf{Layer 3 (Selected)} & 2.06 & 1       & 55.17                                  & \textbf{55.05}               & \textbf{54.40}                 & \textbf{53.86}               & \textbf{52.14}               \\ \bottomrule
\end{tabular}}
\caption{\textbf{Layer Selection Experiments:} Comparison on the performance between different layers on Qwen-Audio ER task (Full token baseline accuracy: 54.80\%)}
\label{layer_select}
\end{table}
\section{Experiments}
\label{exp}

\subsection{Experiment Setting}
\textbf{Basic Settings: }We use 1$\times$V100 32GB GPU to conduct the task performance experiment. We also use 1$\times$A100 80GB GPU and 1$\times$H100 80GB GPU for long sequence system metric experiment. We choose WavLLM 7B~\cite{hu2024wavllm} and Qwen-Audio 7B~\cite{chu2023qwenaudio} for all the experiments. For each SpeechLM, we choose two \textit{dense tasks} and two \textit{sparse tasks} for experiments. Specifically, both models choose ASR and ST as dense task. For sparse task, we choose Emotion Recognition (ER) and Audio Caption (AC) on Qwen-Audio; ER and SQA on WavLLM. The full details of the dataset information and the evaluation metrics can be found in Table \ref{tasks}.

\noindent\textbf{System Metrics:}
We use Theoretical FLOPs, Real Time Factor (RTF), Pre-filling and Decoding Latency (seconds per sentence), and Throughput (tokens per second) to measure the efficiency of our method under different token reduction rates. We calculate the RTF by:
\begin{equation}
\text{RTF} = \frac{T_\text{Pre-filling}+T_\text{Decoding}}{T_\text{audio}} 
\end{equation}
Where $T_\text{Pre-filling}$ and $T_\text{Decoding}$ represents the pre-filling and decoding latency, $T_\text{audio}$ represents the audio length (second per sentence).

\subsection{Results and Discussion}
\label{result}
In this section, we compare our method with other SOTA methods. Then, we demonstrate the impact of token reduction on system metrics. For the full experiments results, please refer to Appendix \ref{fullexp}.


\noindent\textbf{Baselines: }We selected several token reduction methods as our baselines. FastV~\cite{chen2024imageworth12tokens} is a token eviction method based on attention scores for VLM. A-ToMe~\cite{li2023accelerating} incorporates pair-wise merging techniques on the Transducer Model for ASR. We also test two other baselines method which randomly merge or evict tokens as the additional reference. Additionally, we applied our layer selection method to FastV and the two other random baselines since they do not have a clear layer selection strategy for speech tasks. Randomly choosing layers for these methods could result in completely failed decoding. Lastly, we evaluate the performance of the KV cache eviction method (H$_2$O)~\cite{zhang2023h2o} on SpeechLMs for reference. However, this method is primarily designed to accelerate multi-round generation, focusing on a different set of challenges and applications compared to our work.

\begin{table*}[]
\centering
\begin{tabular}{@{}c|ccccc|ccccc@{}}
\toprule
               & \multicolumn{5}{c|}{ASR (WER\% $\downarrow$)}                                                & \multicolumn{5}{c}{ER (ACC\% $\uparrow$)}                                                    \\ \midrule \rowcolor[HTML]{C0C0C0} 
Full Token Baseline                                                                                    & \multicolumn{5}{c|}{\cellcolor[HTML]{C0C0C0}2.25}                             & \multicolumn{5}{c}{\cellcolor[HTML]{C0C0C0}72.80}                                  \\ \midrule
FLOPs Reduce   & 10\%           & 20\%           & 30\%           & 40\%            & 50\%            & 10\%            & 20\%           & 30\%            & 40\%            & 50\%            \\ \midrule
Average Merge  & 2.25          & 2.53          & 3.92          & 12.78          & 93.14          & 72.52          & 72.28         & 73             & 71.95          & 72.84          \\
Weighted Merge & \textbf{2.25} & \textbf{2.44} & \textbf{3.25} & \textbf{10.51} & \textbf{90.24} & \textbf{73.16} & \textbf{72.6} & \textbf{73.73} & \textbf{73.65} & \textbf{73.65} \\ \bottomrule
\end{tabular}
\caption{\textbf{Average Merge vs. Weighted Merge. }The effectiveness of weighted merge method on WavLLM for both Dense and Sparse Tasks}
\label{weighted_exp}
\end{table*}

\begin{table*}[]
\centering
\resizebox{\textwidth}{!}{\begin{tabular}{@{}c|ccccc|ccccc@{}}
\toprule
               & \multicolumn{5}{c|}{ASR (WER\% $\downarrow$)}                                                & \multicolumn{5}{c}{ST (BLEU $\uparrow$)}                                                    \\ \midrule \rowcolor[HTML]{C0C0C0} 
Full Token Baseline                                                                                    & \multicolumn{5}{c|}{\cellcolor[HTML]{C0C0C0}2.25}                             & \multicolumn{5}{c}{\cellcolor[HTML]{C0C0C0}21.56}                                  \\ \midrule
FLOPs Reduce   & 10\%           & 20\%           & 30\%           & 40\%            & 50\%            & 10\%            & 20\%           & 30\%            & 40\%            & 50\%            \\\midrule

\multicolumn{1}{l|} {Weighted Merge}  & \textbf{2.25}          & \textbf{2.44}          & 3.25          & 10.51          & 93.14          & 20.94          & 20.03    & 18.41        & 14.45       & 8.74      \\
\multicolumn{1}{l|} {Weighted Merge + Constant Schedule} & 2.27    & 2.49       & \textbf{2.48}        & \textbf{2.96}     & \textbf{4.73}          & \textbf{21.47}          & \textbf{20.72}    & \textbf{19.81}        & \textbf{18.54}       & \textbf{17.45}      \\
\multicolumn{1}{l|} {Weighted Merge + Decay Schedule} & 2.27    & 2.57       & 2.74         & 3.53     & 6.09          & 20.92          & 20.59    & 19.66        & 18.06       & 16.40      \\

\bottomrule
\end{tabular}}
\caption{The effectiveness of scheduler on WavLLM Dense tasks (ASR and ST)}
\label{schedule}
\end{table*}

\noindent\textbf{Efficient Inference for Dense Tasks:} We selected ASR and ST as the dense tasks in SpeechLM. As shown in Table \ref{dense}, our method demonstrates a significantly better efficiency-performance trade-off compared to other token reduction methods. Notably, for the ASR task, we maintain only approximately 0.7\% WER degradation up to a 40\% FLOPs reduction ratio. Furthermore, we significantly improve upon the previous audio efficient inference baseline, A-ToMe, reducing the WER from 50.46\% to 4.73\% at a 50\% FLOPs reduction rate. For the ST task, our method also maintain the best efficiency-performance trade-off with only approximate 4 BLEU score degradation on 50\% FLOPs Reduce Rate.

\noindent\textbf{Efficient Inference for Sparse Tasks:} For the \textit{Sparse Task} result in Table \ref{sparse}, our method not only surpasses most of the token reduction methods but also improves the original full token baseline from 67.6\% to 68.7\% accuracy for SQA and from 72.8\% to 73.65\% accuracy for ER. These experimental results demonstrate that sparse tasks can be enhanced by the token reduction method, which helps the model ignore redundant audio tokens in a more effective manner. 

\noindent\textbf{Computational Cost Analysis:} We analyze our token reduction method across various system metrics and demonstrate efficiency improvements at a 50\% token reduction rate. The results in Table \ref{computation} show that we achieved a \textbf{1.84x} increase in decoding throughput (from 3.10 tokens/s to 5.72 tokens/s) under A100 GPU and a \textbf{1.44x} throughput under H100 GPU. Further, our method can also decrease both pre-filling and decoding latency at about 4\% and 50\%, respectively.

\noindent\textbf{Memory Saving Analysis:} For memory efficiency in batch decoding settings, as shown in Table \ref{memory}, our system can achieve approximately a \textbf{7x} increase in batch size after a 50\% token reduction in practical deployment. These improvements demonstrate the significant potential of our token reduction method in enhancing both computational and memory efficiency for large-scale applications.

\subsection{Ablation Study}

\noindent\textbf{Effectiveness of Layer Selection: }We analyze the effectiveness of our TE-based layer selection method in Table \ref{layer_select} as an ablation study. Several operation layers before layer 15 were selected to analyze the relationship between the TE and their actual performance. The results indicate that selecting the operational layer based on the TE rank (layer 3) can achieve the best performance on the ER task at most of the time. While the rank of TE may not be strictly proportional to the actual performance, in our study, TE serves as a theoretical reference for layer selection. A more comprehensive study on layer selection for token reduction is left for future research.


\noindent\textbf{Effectiveness of Weighted Merge:} Table \ref{weighted_exp} clearly illustrates the effectiveness of the weighted merge method. Compared to the normal average merge used in ToMe~\cite{bolya2023token} and A-ToMe~\cite{li2023accelerating}, our weighted merge algorithm consistently improves both ASR and ER in all the 10\% to 50\% FLOPs reduction ratio.

\noindent\textbf{Effectiveness of Scheduling:}
\label{appendix_schedule}
For the dense tasks ASR and ST, we utilize the decay or constant scheduler to smoothly merge audio tokens which can prevent aggressive token dropping. As shown in Table \ref{schedule}, layer scheduler can greatly improve the performance of the dense task when the token reduction rate is very high. However, due to multiple operations across many layers, the pre-filling latency will increase. Therefore, a more careful design of the overall strategies is needed in the future to better manage the trade-off between performance and efficiency.

\section{Conclusion}
In this study, we propose \textbf{FastAdaSP}, an efficient inference framework that incorporates multiple stages in SpeechLMs. This preliminary study explores token reduction methods for SpeechLMs. We investigated various properties of different types of SpeechLM tasks and proposed novel methods for both dense and sparse tasks. Our method achieved a \textbf{1.84x} throughput increase with \textbf{7x} memory efficiency, setting a new benchmark for the efficiency-performance trade-off across various tasks.


\section*{Acknowledgments}
Experiments of this work used the Bridges2 system at PSC  and Delta system at NCSA through allocations CIS210014 and IRI120008P from the Advanced Cyber infrastructure Coordination Ecosystem: Services \& Support (ACCESS) program, supported by National Science 
Foundation grants \#2138259, \#2138286, \#2138307, \#2137603, and \#2138296.

\clearpage
\bibliography{main}

\newpage
\appendix

\clearpage

\begin{table*}[htbp]
\centering
\resizebox{\textwidth}{!}{\begin{tabular}{@{}c|ccccc|ccccc@{}}
\toprule
                                                                                            & \multicolumn{5}{c|}{ASR (WER\% $\downarrow$)}                                              & \multicolumn{5}{c}{ST (BLEU $\uparrow$)}                                                       \\ \midrule
\rowcolor[HTML]{C0C0C0} 
Full Token Baseline                                                                                    & \multicolumn{5}{c|}{\cellcolor[HTML]{C0C0C0}2.21}                             & \multicolumn{5}{c}{\cellcolor[HTML]{C0C0C0}41.46}                                  \\ \midrule
FLOPs Reduce                                                                                & 10\%          & 20\%          & 30\%          & 40\%          & 50\%          & 10\%           & 20\%           & 30\%           & 40\%           & 50\%           \\ \midrule
Random Merge                                                                                & 2.43          & 3.39          & 8.21          & 27.53           & 169.96         & 40.63          & 39.35          & 37.01          & 32.39          & 24.3           \\
Random Evict                                                                                & 5.70          & 21.42          & 61.04          & 184.59          & 342.88        & 38.39          & 28.22          & 14.98          & 6.29           & -           \\
A-ToMe~\cite{li2023accelerating}                                                                                      & 2.20          & 3.26          & 13.91          & 71.56         & 273.49         & 41.24          & 39.87          & 36.52          & 25.35          & 8.64           \\
FastV~\cite{chen2024imageworth12tokens}                                                                                & 12.54          & 54.40          & 110.42         & 179.58          & 258.78        & 41.12          & 40.31          & 38.45          & 34.74          & 27.14          \\ \midrule
\textbf{\begin{tabular}[c]{@{}c@{}}FastAdaSP-Dense\\ {\small Decay Schedule}\end{tabular}}    & \textbf{2.19} & 2.23          & 2.51          & 4.37          & \textbf{15.24 }         & 41.41          & 41.05         & 40.51          & 39.02          & 35.79           \\
\textbf{\begin{tabular}[c]{@{}c@{}}FastAdaSP-Dense\\ {\small Constant Schedule}\end{tabular}} & 2.22  & \textbf{2.21} & \textbf{2.30} & \textbf{3.57} & 16.01  & \textbf{41.47} & \textbf{41.30} & \textbf{40.83} & \textbf{39.81} & \textbf{37.04} \\ \bottomrule
\end{tabular}}
\caption{Comparison between FastAdaSP with other token reduction methods on Qwen-Audio \textbf{dense tasks}}
\label{qwen-dense}
\end{table*}

\begin{table*}[ht]
\centering
\resizebox{\textwidth}{!}{\begin{tabular}{@{}c|ccccc|ccccc@{}}
\toprule
                           & \multicolumn{5}{c|}{ER (ACC\% $\uparrow$)}                                                    & \multicolumn{5}{c}{AC (CIDEr $\uparrow$ | SPICE $\uparrow$ | SPIDEr $\uparrow$)}                                                    \\ \midrule
\rowcolor[HTML]{C0C0C0} 
Full Token Baseline                   & \multicolumn{5}{c|}{\cellcolor[HTML]{C0C0C0}54.80}                                  & \multicolumn{5}{c}{\cellcolor[HTML]{C0C0C0}0.45 | 0.13 | 0.29}                                  \\ \midrule
FLOPs Reduce               & 10\%           & 20\%           & 30\%           & 40\%           & 50\%           & 10\%           & 20\%           & 30\%           & 40\%           & 50\%          \\ \midrule
Random Merge               & 51.80          & 48.00          & 43.80          & 39.20          & 32.30          & 0.44 | 0.13 | 0.29           & 0.43 | 0.13 | 0.28     & 0.41 | 0.13 | 0.27        & 0.41 | 0.12 | 0.26           & 0.38 | 0.12 | 0.25                  \\
Random Evict               & 52.80          & 48.20          & 42.00          & 34.61          & 23.14          & 0.43 | 0.13 | 0.28          & 0.42 | 0.13 | 0.27          & 0.38 | 0.12 | 0.25          & 0.31 | 0.10 | 0.20           & 0.12 | 0.07 | 0.14          \\
A-ToMe~\cite{li2023accelerating}                     & 54.91          & 54.70 & 54.20           & \textbf{53.90}          & 51.60          & 0.44 | 0.13 
| 0.29          & 0.44 | 0.13 | 0.29          & 0.43 | 0.13 | 0.28           & 0.41 | 0.13 | 0.27          & 0.39 | 0.12 | 0.28          \\
FastV~\cite{chen2024imageworth12tokens}               & 54.80          & 53.80          & 53.50          & 52.10          & 50.38          & 0.44 | 0.13 | 0.29          & \textbf{0.45} | \textbf{0.13} | \textbf{0.29}         & 0.45 | 0.13 | 0.29 & 0.44 | 0.13 | 0.28           & 0.43 | 0.13 | 0.28         \\ \midrule
\rowcolor[HTML]{FFFFFF} 
\textbf{FastAdaSP-Sparse} & \textbf{55.17} & \textbf{55.05}           & \textbf{54.40} & 53.86 & \textbf{52.14} & \textbf{0.45} | \textbf{0.13} | \textbf{0.29} & 0.44 | 0.13 | 0.29 & \textbf{0.45} | \textbf{0.13} | \textbf{0.29}          & \textbf{0.44} | \textbf{0.13} | \textbf{0.28} & \textbf{0.43} | \textbf{0.13} | \textbf{0.28} \\ \bottomrule
\end{tabular}}
\caption{Comparison between FastAdaSP with other token reduction methods on Qwen-Audio \textbf{sparse task}}
\label{qwen-sparse}
\end{table*}

\begin{table*}[ht]
\centering
\resizebox{\textwidth}{!}{\begin{tabular}{@{}c|cccccc@{}}
\toprule
Token Reduce \% & SUM (ROUGE-L $\uparrow$) & FLOPs Reduction \% $\uparrow$ & Real Time Factor $\downarrow$ & Pre-filling Latency (s) $\downarrow$& Decoding Latency (s) $\downarrow$& Throughput (token/s) $\uparrow$\\ \midrule
 Full Token Baseline     & 16.20         & 0.00                        & 0.091     &   0.51   & 5.09                      & 15.57                         \\  \midrule
10        & 16.63       & 9.54                     & 0.090     &    0.51   & 4.92                      & 16.00                        \\
20        & 16.27       & 19.05                    & 0.087     &    0.51   & 4.74                      & 16.02                        \\
30        & 16.29       & 28.46                    & 0.083     &    0.51   & 4.52                      & 16.05                       \\
40        & 15.29       & 37.66                    & 0.083     &    0.51   & 4.51                      & 16.62                        \\
50        & 15.10       & 46.97                    & 0.078     &    0.51   & 4.20                      & 16.87                       \\ \bottomrule
\end{tabular}}
\caption{\textbf{Computational cost experiments on Real Dataset.} Inference result of 100 How2 test set samples around 60s on WavLLM using one V100 32GB GPU }
\label{how2}
\end{table*}
\begin{table*}[ht]
\centering
\resizebox{\textwidth}{!}{\begin{tabular}{@{}c|c|cccccc@{}}
\toprule
Beam Size  &  Audio Length (s) & Token Reduce \%  & FLOPs Reduction \% $\uparrow$ & Real Time Factor $\downarrow$ & Pre-filling Latency (s) $\downarrow$& Decoding Latency (s) $\downarrow$& Throughput (token/s) $\uparrow$\\ \midrule

\multirow{2}{*}{1}       & \multirow{2}{*}{120}   & Full Token  & 0.00                    & 0.054     &    0.79   & 5.75                      & 12.86                 \\
        &    &  50         & 48.62                   & 0.044     &    0.77   & 4.57                      & 13.57 (1.05x)                \\\midrule
\multirow{2}{*}{5}        & \multirow{2}{*}{120}   & Full Token  & 0.00                     & 0.137     &    3.11   & 13.32                      & 5.48                 \\
        &    &  50         & 48.40                    & 0.092     &    3.09   & 8.01                      & 8.87 (1.61x)    \\ \midrule 
 \multirow{2}{*}{1}        & \multirow{2}{*}{240}   &  Full Token & 0.00                    & 0.044     &    1.70   & 8.90                      & 8.09                   \\
        &    &  50         & 49.21                     & 0.036     &   1.59   & 7.02                      & 9.69 (1.20x)           \\ \midrule
 \multirow{2}{*}{5}        & \multirow{2}{*}{240}   &  Full Token & 0.00                    & 0.126     &    6.72   & 23.55                      & 3.10                   \\
        &    &  50         & 49.21                    & 0.077     &    6.48   & 11.89                      & 5.72 \textbf{(1.84x)}        \\\bottomrule
\end{tabular}}
\caption{\textbf{Long Sequence Computational cost experiments on A100.} Long sequence audio samples (120s and 240s) input on WavLLM using one A100 80GB GPU }
\label{a100}
\end{table*}
\begin{table*}[!htbp]
\centering
\resizebox{\textwidth}{!}{\begin{tabular}{@{}c|c|cccccc@{}}
\toprule
Beam Size  &  Audio Length (s) & Token Reduce \%  & FLOPs Reduction \% $\uparrow$ & Real Time Factor $\downarrow$ & Pre-filling Latency (s) $\downarrow$& Decoding Latency (s) $\downarrow$& Throughput (token/s) $\uparrow$\\ \midrule

\multirow{2}{*}{1}       & \multirow{2}{*}{120}   & Full Token  & 0.00                    & 0.027     &    0.26   & 3.00                      & 24.63                 \\
        &    &  50         & 48.62                   & 0.023     &    0.26   & 2.52                      & 24.73 (1.01x)                 \\\midrule
 \multirow{2}{*}{5}        & \multirow{2}{*}{120}   & Full Token  & 0.00                     & 0.043     &    0.48   & 4.73                      & 15.44                 \\
        &    &  50         & 48.40                    & 0.032     &    0.46   & 3.44                      & 20.66 (1.34x)    \\ \midrule
 \multirow{2}{*}{1}        & \multirow{2}{*}{240}   &  Full Token & 0.00                    & 0.020&    0.43& 4.29& 16.70\\
        &    &  50         & 49.21                     & 0.019&   0.39& 4.06& 16.75 (1.00x)\\ \midrule
 \multirow{2}{*}{5}        & \multirow{2}{*}{240}   &  Full Token & 0.00                    & 0.039&    1.13& 8.39& 8.70\\
        &    &  50         & 49.21                    & 0.026&    0.96& 5.42& 12.55 \textbf{(1.44x)}\\\bottomrule
\end{tabular}}
\caption{\textbf{Long Sequence Computational cost experiments on H100.} Long sequence audio samples (120s and 240s) input on WavLLM using one H100 80GB GPU }
\label{h100}
\end{table*}

\section{Appendix}



\subsection{Full Experiments Results}
\label{fullexp}
We also conduct the performance experiments on Qwen-Audio for both dense and sparse tasks and compare the baseline methods with our method. For the dense tasks ASR and ST, the results are presented in Table \ref{qwen-dense}, demonstrating the effectiveness of our scheduling weighted token merge methods on another SpeechLM. The results for the sparse tasks ER and AC are shown in Table \ref{qwen-sparse}, which suggest our sparse setting method also performs well. These results on Qwen-Audio shows the effectiveness and generalization of our method across different SpeechLM.

Additionally, for the computation cost experiment, we also evaluated the Speech Summarization task on WavLLM using a subset of the How2 test set \cite{sanabria2018how2largescaledatasetmultimodal}. As shown in Table \ref{how2}, our method can effectively reduce the computation cost on a real dataset. 

Further, we use one A100 80G GPU and one H100 80G GPU to conduct the long sequence experiments, which is shown in Table \ref{a100} and Table \ref{h100}. The results indicate that increasing the audio length and beam size makes the acceleration of our method more noticeable.

\subsection{Computation Reduction Theoretical Analysis}
\label{flops}
To analyze the computation reduction effect of our method, we use the theoretical FLOPs reduction rate. For simplicity, we just analysis the effective theoretical FLOPs reduction based on the token reduction rate and input sequence length on one layer. In the real situation, we can use the same methods to analyse all the decoder layers. Given the input sequence length $n$, the hidden dimension $d$ and the Feed Forward Layer hidden dimension $m$. We can define the theoretical FLOPs in one transformer decoder layer as:
\begin{equation}
    \text{FLOPs} = 2n^2d + 4nd^2 + 2ndm.
\end{equation}

\begin{algorithm}[ht]
\caption{Weighted Token Merge Algorithm}\label{FastAdaSP}
\begin{algorithmic}[1]
\Procedure{FastAdaSP}{$A\in\mathbb{R}^{L \times D}, M\in\mathbb{R}^{T}, W_\text{merge}\in\mathbb{R}^{L}$}

\State $i \gets 1$  \Comment{Index}
\State $H \gets [\ ]$ \Comment{New hidden states}
\While{$i \leq L$}
\State $S \gets \varnothing$ \Comment{Initialize merge cluster}
\State $\mathbf{h} \gets \omega_i\mathbf{a}_i$ 
\State $t \gets \omega_i$
\\
\State \# Form the merge cluster
\While{$i \in M$}
\State $S \gets S\ \cup\ \{i\}$ 
\State $i \gets i + 1$
\EndWhile
\\
\State \# Perform Weighted Sum in Cluster
\For{$j\ \textbf{in}\ S$}
\State $\mathbf{h} \gets \mathbf{h} + \omega_j\mathbf{a}_j$ 
\State $t \gets t + \omega_j$
\EndFor

\State $\mathbf{h} \gets \mathbf{h}\ /\ t$ 
\State $H \gets \append(H, \mathbf{h})$
\State $i \gets i + 1$
        
\EndWhile
\Output 
$ :H \in\mathbb{R}^{N \times D}$
\EndOutput

\EndProcedure
\end{algorithmic}
\end{algorithm}
Where the first term represents the attention operation in equation \ref{attn}; The second term represents the calculation of query, key, value and output tensors; The third term represents the calculation of the operation in Feed Forward Layer. Given the reduction ratio $k$, after the token reduction, we obtain the reduced sequence length $\hat{n} = n(1-k)$. Then the theoretical FLOPs reduction rate at the next layer can be calculated as:
\begin{align*}
    \text{Rate} &= 1-\frac{2\hat{n}^2d + 4\hat{n}d^2 + 2\hat{n}dm}{2n^2d + 4nd^2 + 2ndm} \\
    &= 1 - \frac{2(1-k)^2n^2d+nd(1-k)(4d + 2m)}{2n^2d + nd(4d + 2m)} \\
    &= k + \frac{(k-k^2)}{1 + \frac{(2d + m)}{n}} \propto n.
\end{align*}

As a result, the longer the input sequence length, the higher the FLOPs reduction rate that can be achieved. As demonstrated in \ref{fullexp} long sequence speed test, the acceleration is more pronounced for a 240-second audio sample compared to a 120-second audio sample.

This theoretical computation cost analysis suggests that our method will result in greater computational reduction for longer audio sequence input, highlighting the effectiveness of this technique in real world applications where the input audio is often very long.

\subsection{FastAdaSP: Algorithm Details}
\label{algo}
Here we show the full implementation details of the \textbf{FastAdaSP} algorithm, which was brifely mentioned in Section \ref{method}. Given the audio feature sequence $A = (\mathbf{a}_{i}\in\mathbb{R}^D|i = 1, ..., L)$, the merge index list $M = (m_{i}\in\mathbb{R}|i = 1, ..., T)$ and merge weights $W_\text{merge}=(\omega_{i}\in\mathbb{R}|i = 1, ..., L)$. Then we can use Algorithm \ref{FastAdaSP} to obtain the merged audio feature sequence $H = (\mathbf{h}_{i}\in\mathbb{R}^D|i = 1, ..., N)$, where $N$ is the length of the merged audio feature sequence. 

Additionally, if there are $B$ batches in the hidden states, we currently need to perform the algorithm $B$ times to reduce the audio tokens for each audio sequence separately. In the future, this process may be improved by executing the algorithm for each batch in parallel.

\subsection{Derivation of Transfer Entropy}
\label{te_derivation}
In this section, we recall the derivation of transfer entropy from~\cite{lin2024mlpgoodtransformerlearner}. We also did a slight modification on the final definition based on our settings. As mentioned in section \ref{layer}, given $F \in \mathbb{R}^{L \times D}$ as the feature output after attention block, the entropy was defined as:
\begin{equation}
\label{eq10}
    H(F) = -\int p(f) \log p(f) \, df, \ f \in F.
\end{equation}

Following the ~\cite{lin2024mlpgoodtransformerlearner, sirignano2019meanfieldanalysisneural}, we regard the feature $F$'s probability distribution as a Gaussian distribution $F \sim \mathcal{N}(\mu, \sigma^2)$. Therefore, the equation \ref{eq10} can be derived into:

\begin{small}
\begin{align*}
H(F) &= -\mathbb{E}[\log \mathcal{N}(\mu, \sigma^2)] \\
     &= -\mathbb{E}\left[\log \left[(2\pi\sigma^2)^{-\frac{1}{2}} \exp\left(-\frac{1}{2\sigma^2}(f - \mu)^2\right)\right]\right] \\
     &= \log(\sigma) + \frac{1}{2} \log(2\pi) + \frac{1}{2}
\end{align*}    
\end{small}

Where $\sigma_i$ is the standard deviation of $i$-th hidden state in $F$. The $H(F)$ is proportional to the $\log(\sigma)$ since $\frac{1}{2} \log(2\pi) + \frac{1}{2}$ is constant term. Thus we could get the equation \ref{eq_te} in Sec \ref{layer}.

\subsection{Applications in the Real World and Future Perspective}


In this study, we propose a efficient inference framework which designed for audio modality reduction in Multitask SpeechLM. In the context of long audio sequences, it is observed that only a small part of tokens carries critical information, while others may be not relevant (e.g. periods of noisy or blank audio). Our proposed plug-and-play methodology aims to efficiently identify and prioritize significant audio tokens during the pre-filling stage, which can offers substantial benefits for long-form audio comprehension.

In addition, in practical deployments of SpeechLM products, batch decoding is often a necessity, with batch sizes potentially reaching up to 128 or more. Within these batch decoding settings, our proposed methods are designed to reduce the memory footprint associated with many long audio inputs while simultaneously accelerating the decoding process. This optimization is crucial for enhancing the efficiency and scalability of SpeechLM systems in real world applications. 

In the future, we may extend the current efficient inference framework to multi-round decoding scenarios, which can handle the dense task and sparse task at the same time. This improvement will make the whole system more applicable to real world use cases. Moving forward, this pioneering study on audio token reduction techniques in Multimodal Large Language Models (MLLM) paves the way for future research to explore the general behavior of audio and other modalities such as vision. The next stage of this study is to investigate the unified methodology to accelerate both audio and vision modalities simultaneously in Audio-Visual LLMs (e.g., video-SALMONN~\cite{sun2024videosalmonnspeechenhancedaudiovisuallarge}), which enable more efficient inference for long video understanding.




\end{document}